# Computer Supported Collaborative Research

VITA HINZE-HOARE,
UNIVERSITY OF SOUTHAMPTON

___________________________________________________________________

It is suggested that a new area of CSCR (Computer Supported Collaborative Research) is distinguished from CSCW (Computer Supported Collaborative Work) and CSCL (Computer Supported Collaborative Learning) and that the demarcation between the three areas could do with greater clarification and prescription.

Although the areas of Human Computer Interaction (HCI), CSCW, and CSCL are now relatively well established, the related field of Computer Supported Collaborative Research (CSCR) is new and little understood. An analysis of the principles and issues behind CSCR is undertaken with a view to determining precisely its nature and scope and to delineate it clearly from CSCW and CSCL. This determination is such that it is generally applicable to the building, design and evaluation of collaborative research environments.

A particular instance of the CSCR domain is then examined in order to determine the requirements of a collaborative research environment for students and supervisors (CRESS).

Categories and Subject Descriptors: H.5.2 (Information Interfaces and Presentation):

User Interfaces - User centred design

General Terms: Computer Supported Collaborative Research CSCR, Human Factors

___________________________________________________________________

## 1. INTRODUCTION

The field of Computer Supported Collaborative Research (CSCR) is not yet established. There is very little in the literature concerning the significant issues which arise in the design of a support system for collaborative researchers to enable them to work together effectively from a distance. Much has been written about the twin related fields of Computer Supported Collaborative Work (CSCW) and Computer Supported Collaborative Learning (CSCL) which have been the subject of intense interest in the HCI research community during recent years. CSCR on the other hand has arisen from within these fields recently.

The research was supported by the University of Southampton and by HEFCE
Authors' addresses: Vita Hinze-Hoare, School of Electronics and Computer Science, University of Southampton, Southampton, SO17 1BJ.

The domain of CSCR needs to be strictly defined and determined together with a clear differentiation between this domain and the domains of CSCW and CSCL. Once this determination has been made it will define the characteristics of all collaborative research environments and interfaces. At the present time there is no clear definition of these domains and the purpose of this paper is t o bring these into sharper focus to enable a clearer understanding of what is required when research environments are being constructed. The definition of CSCR provided here will be applicable to all collaborative research environments and is presented as the specification which all such environments should meet. Furthermore by association the definition of CSCW and CSCL are also presented.

## 2. BACKGROUND

The History of HCI shows a lack of coherent development. There is no agreement as to what HCI is, should be, or does. Diaper [2005] The discipline is becoming increasingly fragmented to the point where it is difficult to establish consensus in the field. This fragmentation of discipline of HCI is already so extensive according to Diaper that it is hard to even characterise the method of approach

.

Much the same is true of CSCW and CSCL. Although these have been the subject of extensive research for a number of years there is still no accepted definition of either. "*This lack of agreement highlights the necessity for the development of a general systems model, both in the general HCI approach and in the specific collaborative approach*" Diaper [2005]

The split between CSCW and CSCL has grown wider in response to the recognition that the learning process is distinct from the working process and the former is more intensively understood through new theories of pedagogy and education. Furthermore, it is recognised that the distinction between learning and research leads to its own requirements and issues for a collaborative framework.

The relationships between CSCW, CSCL and CSCR are determined by the differences between work, learning and research. Learning is a specific type of work and research is a specific type of learning. The process of research is a learning process but one which is highly refined and involves learning in a particular way.

*"Research is the creation of new knowledge".[1]*
*"Research encompasses activities that increase the sum of human knowledge" [OECD Definition].[2]*

Thus, the nature of research means that the body of knowledge cannot be taught but must be discovered. The research process is an extension of the normal learning and teaching process. As such it can be further argued that research supported by computer collaboration is an extension of CSCL. See Figure 1.

It is becoming apparent that CSCL is part of CSCW but is constraint by additional needs of pedagogical theories. In addition, it is also becoming apparent that CSCR is part of CSCL but is constrained by the additional requirements of research. Research is understood to be a highly specialised and refined learning process that takes place without the presence of a teaching environment. This requires new mechanisms of independent knowledge acquisition and the support of these activities with new techniques and tools.

---

[1] www.universities-scotland.ac.uk/Facts%20and%20Figures/Research.pdf

[2] www.jcu.edu.au/office/research_office/researchdef.html

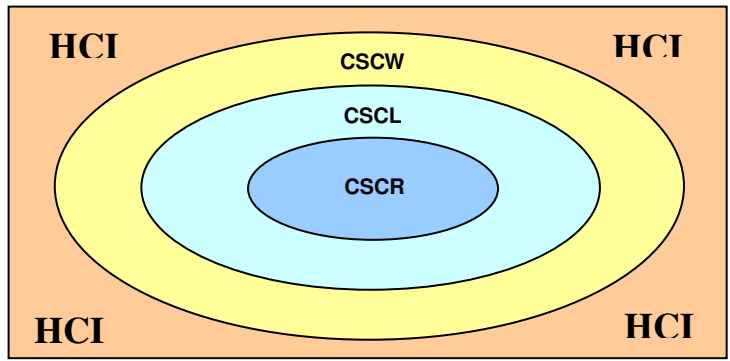

Figure 1 Collaborative Domains within HCI

3. DELINEATING CSCW, CSCL AND CSCR

It is contended that CSCW, CSCL and CSCR are domains within the HCI field and consequently suffer from the same lack of distinctive definition that bedevils HCI. The purpose of this section will be to propose specific and clear definitions of CSCW and CSCL and show that neither is sufficient to support the research domain CSCR. These are now addressed in turn.

3.1 Determining the CSCW Domain

When HCI is applied to the specific area of Collaborative Work it is commonly known as CSCW. This requires an analysis of collaboration in the workplace on top of HCI principles. The new features of collaboration and the way in which this is analysed and structured form the basis of this section. Various definitions of CSCW have been proposed but none of them have sought to differentiate the domain from CSCL and CSCR. Nor have they sought to express all three in a single constructive framework.

Hawryszkiewycz, I. [1994] has proposed a definition of CSCW based upon semantic elements as follows:

- Artefacts [files, reports, documents, policies etc]
- Actors [a person in the organisation, each person can play many parts]
- Tasks [well defined business functions]
- Activities [the processes for interactions between artefacts]
- Environments [provide the supportive structures]

These elements are combined to model the system using diagrams which are similar to Systems Analysis diagrams. While this may be helpful from an operational point of view it does not seek to provide a clear differentiation with CSCL or CSCR and therefore its usefulness as a conceptual framework is limited.

Muller and Wu [2005] have proposed an alternative view. They propose that CSCW should be viewed as structured around five landmark entities :
- Documents [including drafts]
- Dates and Calendars [particularly start and end dates]
- Events [including the "kick off meeting"]
- Roles and persons
- Systems and databases

Again, this is open to the same criticism that it is operational rather than conceptual and therefore limited in the framework which it affords. Each of these entities may play a role in CSCW, CSCL and CSCR but they are not sufficient to distinguish the domains.

Carroll et al [2006] has approached CSCW from a more primitive standpoint. They ask the fundamental question *"What do Collaborators need to share in order to work together effectively"*. They derive four design requirements for effective CSCW:
- public display of shared information,

- integration of data into community metaphors to facilitate analysis,
- aggregation of individual contributions into collective overviews to evoke trust and commitment
- contrast of individual capabilities and roles to invite collaborators to perform beyond themselves

These four primitives go much further than Hawryszkiewycz, I. [1994] and Muller and Wu [2005] in that they provide some degree of determination of the CSCW domain. However, there is no attempt to link this framework of primitives into a larger framework to encompass the CSCL and CSCR domains.

The inadequacy of these prior definitions has brought us to the point were we can see the need to construct a framework which tightly delineates the requirements of a CSCW domain and integrates it into a larger framework which includes the CSCL and CSCR domains. It will be contended that each domain is determined by a set of working spaces which specify their content and link them together in a clear and systematic way.

The wider domain of CSCW will now be specified here as containing a number of distinct spaces which contain specific activities. CSCW is constructed from Communication, Identification, Scheduling, Sharing, Product and Administration spaces.

**Communication Space**

The first space that is essential to CSCW deals with communication and the interchange of ideas. The domain will be expected to include real time collaboration as well as asynchronous communication. The use of a whiteboard and video/audio channels are primarily all real time communication devices, while bulletin boards, email and message centres are asynchronous devices. Watts and Reeves [2005] have pointed out that email lacks social sensitivity and can be detrimental to communication by fostering misunderstanding.

However the communication space is not limited to these devices. Karam and Schraefel [2005] have added another dimension by examining the role of gestures as an additional communication device. They examined gestures with reference to HCI in order to see if this provides the necessary richness for effective collaborative communication. They provide a literature review of over 40 years of gesture based interactions which they then categorize into a taxonomy of gestures denoted by four key elements: Gesture styles, Enabling technologies, Application domain and System response.

Karam and Schraefel conclude that gestures are a natural, novel and improved mode of interaction. However we suggest that it will take some time before they are incorporated as a standard feature into Microsoft Windows in the same way that say speech has been incorporated. They have demonstrated that there is a vast range of research in this area but very little application as yet. Although gestures are a part of communication space, it still needs to be assessed how important they are for effective collaborative work, and it needs to be carefully considered whether we need to include gestures in the collaborative interface.

Additional communication methods have been raised by other researchers. The use of digital backchannels in the context of an academic conference have been discussed by McCarthy and Boyd [2005] . This involves members of the audience communicating with each other via laptops during a speaker's presentation via chat channels which are opened up to allow all participants to discuss the presentation thereby adding information to what is being presented.

**Identification Space**

The second essential space to the CSCW domain involves the process of identification and tracking. In order for communication to be coherent it is essential that the participants are clearly identified in one way or another. This will include "anonymous identification" where pseudonyms may be used.

Juby and De Roure [2002] have argued that real time collaboration requires more than just audio, video and data sharing, and have proposed two specific enhancements to provide a richness of interaction that is required for proper collaboration which are "speaker identification and participant tracking for the automatic generation of dynamically updated attendance lists". The essential nature of identification is conceded and will be incorporated in the definition of the CSCW domain.

It may be objected to by some that identification is not required for clear communication as it is possible to exchange ideas without full knowledge of the source. Indeed anonymity can enhance communication. Postmes et al [2001] have found that by allowing contributors to remain anonymous throughout their communications they are prepared to interact more, become more vocal participants and show a higher degree of influence within a group. This is often the case when junior members of the team feel intimidated by senior or dominate members. Sassenberg and Postmes [2002] have further concluded that the use of photographs of group members meant that individuality became more important even if incorrect photographs were shown. Spears et al [2002] concur with Postmes that isolation and anonymity in cyberspace produce more social interactions rather than fewer. People can be more outspoken online than they would be in real life which can lead to social repercussions if the anonymity is taken away.

However in any discussion it is essential to be able to verify the source of each statement so as to enable the tracking of ideas. This is possible with pseudonyms or what may be termed as "anonymous identification" which is a necessary minimum for communication. Participants need identification. Careful consideration needs to be given to the role of anonymity in a research environment. Although anonymity promotes greater social interaction Postmes *et al* [2001] this may not be the most important requirement. Even more important may be the need for reliability of information and being able to trace the source of information and establish validity. On the other hand anonymity may be required in the area of peer review to obtain unfettered criticism. The inclusion of some form of identification therefore needs to be included in the CSCW domain but the nature of anonymity and its formalism can be left to a later stage.

### Scheduling Space

The third essential space to the construction will involve the processes of setting up the opportunities for communication. Both synchronous and asynchronous communication require the scheduling of meetings, the setting of deadlines, setting up of conferences [online or otherwise]. A common scheduling facility is required to maintain collaborative interaction. The implications of Rahikainen *et al* [2001] study are that the less able research students need careful and closer monitoring. This will require clear scheduling and task setting interface tools. This is supported by Joiner *et al* [2006] who have shown that students overwhelmingly prefer goal driven scenarios to non-goal driven scenarios. The design of any interface must therefore include consideration of goal setting, target achievement, and personal reward. [Graves and Klawe 1997; Klawe, M. 1999] also support the view that specific goals and target setting are important features.

It is agreed that this research indicates the essential role of scheduling and task setting in order to meet the demands of the collaborative working domain and this has been incorporated into the definition of CSCW.

**Sharing Space**

The fourth essential space is that area which facilitates the interchange of data. The nature of collaboration is by definition determined by a commonality of features which allow this interchange of work to take place. This is where work in progress can be passed between collaborators. This will include groupware, mark-up and revision notes etc. Collaborative research necessitates the exchange of information which may be in multimedia formats such as sound, video, image text etc.

**Product Space**

Artefacts are the expected outcome of the working process and a tally of these needs to be kept and maintained as a record of work done and an indication of progress and the recording of re-iterative work on products.

**Administration Space**

The day to day management of data and the administration of tasks and the maintenance of the interface will require its own area and membership. Bartholome *et al* [2005] conclude that help functions by themselves are not effective. However they are essential components of a larger administration space. In addition facilities to record and replay communications together with instant messaging and assistive agents which provide sophisticated help functions would be part of a necessary administration space for the CSCW domain.

### Comparison with Carroll's CSCW

It has been shown that the newly proposed definition of CSCW requires six determinants whereas Carroll has only specified four determinants which are included in our model.

Table 1 Comparison with Carroll's CSCW determinants

| Proposed New Definition of CSCW | Carroll's Definition of CSCW | Comment |
|---|---|---|
| Communication Space | "*integration of data into community metaphors to facilitate analysis*" | Community metaphors enable each participant to 'speak the same language' and thus to facilitate communication. This will take place in the communication space. |
| Identification Space | "*contrast of individual capabilities and roles to invite collaborators to perform beyond themselves*" | Individual identification and role play are important to contribution. Limitation of identity may have, as we have said, an enhancing effect on contribution. |
| Scheduling Space | | Carroll does not have a correspondence with scheduling space |
| Sharing Space | "*public display of shared information*" | Collaboration is based on sharing information and data. |
| Product Space | "*aggregation of individual contributions into collective overviews to evoke trust and commitment*" | Individual contributions are combined into a collective product which is the purpose of the collaboration. Engendering trust is a by-product essential to CSCW |
| Administration Space | | Carroll does not have a correspondence with administration space |

The two additional spaces are administration and scheduling space ssee table 1. The need for community metaphors as a necessary requisite for common channels of communication is supported by Carroll. The metaphors act as a language which the whole community can use as a transfer of information. This is represented in our model as communication space which carries multiple communication streams and includes Carroll's first determinant.

Our proposal for an Identification space is supported by Carroll in his reference to "*Individual capabilities and roles*" as the second determinant. "Anonymous identification" as defined above has been shown to allow greater participation This is supported by Carroll who makes the point that individuals acting in this environment can often perform "*beyond themselves*". (Postmes *et al* [2001]).

The essential essence of collaborative shared work takes place inside the sharing space of our model. Carroll's third determinant of the public display of shared information concurs.

The product space in our model represents that area where individual contributions are forged into a combined artefact. This is supported by Carroll's fourth determinant "*aggregation of individual* *contributions into collective overviews to evoke trust and commitment"* which recognizes the need for individual work to be combined into a collective product. Collective work producing a single product engenders trust and commitment between collaborators and this is an important by-product of the process which is essential to successful CSCW.

Two additional spaces have been found to be necessary to the successful implementation of CSCW which Carroll has not addressed. The first of these is the scheduling space which is essential for both real time and asynchronous collaborative communication. Without this space it would be impossible to collaborate effectively and would reduce working partners to individuals rather than collaborators. The second space is purely administrative but no less essential than the others. Without administrative procedures and a behind the scenes administrator the CSCW environment would not be able to operate. It might be the case that this role is taken on by the collaborators themselves or by designated individuals but one way or another it needs to be addressed.

### 4.2 Determining the CSCL Domain

CSCL has grown out of CSCW. Table 2 indicates the main differences between CSCW, CSCL and CSCR.

Table 2: Differences between CSCW and CSCL and CSCR

| *CSCW* | *CSCL* | *CSCR* |
| --- | --- | --- |
| *Focuses on communication techniques* | *Focuses on what is being communicated* | *Focuses on new communications* |
| *Used mainly in a business setting* | *Used mainly in an educational setting* | *Used mainly in a research setting* |
| *Purpose is to facilitate group communication and productivity* | *Purpose is to support students in learning together* | *Purpose is to support researchers in working together* |

By definition Computer supported collaborative learning CSCL has four component parts:

- **Learning**- This is seen as an activity that takes place in a wider context than a classroom and involves the everyday social practices of people during which learning occurs and the situation it springs from [Lave and Wenger, 1990]
- **Collaborative learning** – The role of others in the learning process has been highlighted by Vygotsky [1978] and his key concept of the zone of proximal development [ZPD] as the area of overlap between inexperienced and experienced where learning occurs.
- **Computer Supported -** The tools required to provide the environment and the mechanisms for collaboration.
- **Computer supported collaborative learning**. The computer brings a new dimension to the process of learning and introduces a number of new features.

In short CSCL facilitates the learning process through a number of applications including email, computer conferencing, bulletin boards, local area networks, and hypermedia.

A number of researchers have attempted to describe the requirement of a CSCL domain.

It is Bannon's [1989] contention that the best way to regard computers in the CSCL process is as an enabling medium through which partners can organise and accomplish activities. The computer provides a space to work in which others can organise their activities. Although this might be necessary to determine CSCL it is not sufficient and it will be shown that a range of spaces are required.

Lipponen [2002] defines CSCL as being focussed on "*How collaborative learning supported by technology can enhance peer interaction and work in groups and how collaboration and technology* *facilitate sharing and distribution of knowledge and expertise among community members*". Each element in this definition can be considered to be valid within its context. However, this does not go anywhere towards providing a full framework by which CSCL can be fully specified let alone related to the domains of CSCW and CSCR.

Dillenbourg [1999] has characterized CSCL by degrees of symmetry

- Symmetry of action [the extent to which each collaborator has the same range of actions]
- Symmetry of knowledge [the extent to which collaborators possess the same level of skills]
- Symmetry of status [the extent to which collaborators have the same status with respect to their community]

These symmetries are more concerned with modes of interaction rather than a detailed specification of CSCL. They characterize the relationships between the participants rather than determine the requirements for the domain. As such Dillenbourg's work cannot be accepted as a prescription for a determining the essential components of the CSCL domain.

The inadequacy of these prior definitions has brought us to the point were we can see the need to construct a framework which tightly delineates the requirements of a CSCL domain and integrates it into a larger framework which includes the CSCW and CSCR domains. It will be contended that each domain is determined by a set of working spaces which specify their content and link them together in a clear and systematic way.

The wider domain of CSCL will now be specified here as containing a number of distinct spaces which allow the performance of specific activities. It is important to note that since CSCL is a specialized form of CSCW it will contain all of the spaces which determine CSCW together with those additional spaces which are determined by pedagogical constraints. It will be shown that the additional spaces of CSCL are Reflective, Social, Assessment and Supervisor spaces.

### Reflective Space

An important part of learning which has been recognised by pedagogists is the need for internal reflection. [Bruner 1996] This can be both individual and collaborative and could be assisted with the help of an on-line journal (Private and Group) It has been concluded by Dillenbourg [1999] that there is no objective measure of cognitive load. This leads to the suggestion that reflective space will be an important feature of the CSCL domain where personal assessment of progress can be made. More work needs to be done in this area and this might be a suitable topic for further investigation in this research.

### Social Space

Much learning has been shown to arise from interaction with peers and other learners as well as from a didactic intercourse with mentors. [Daniels H. 2001] It is expected that the CSCL system will require additional compensating tools to avoid misunderstanding. Taking account of Watts and Reeves [2005] social links will be incorporated into the CSCL system. The importance of motivation is pointed out by Tapola *et al* [2001]. This is a complex subject to analyse as motivations may come from various sources. However social spaces have been shown to contribute to the motivation of some students and therefore it will be important to consider the inclusion of social space in the CSCL domain. The experience in remote teaching and evaluation of course work using Net Meeting is discussed by Varey [1999]. She claims her experience of collaboration as positive showing student enjoyment of involvement with other students.  In addition new social spaces including Facebook, Digg, Del.icio.us and in the 3D domain Second

Life and its derivatives have all contributed to the establishment of enhanced learning through social networking.

Learning that rakes place in groups has been shown to be more effective than learning individually. This is the basis of social constructivism [Bruner 1996] Even groupings as small as two have shown to be more effective in the learning process. Dillenbourg [1999] conclusion is that it cannot be predicted how social interactions of pairs will affect individual cognitive processes. One therefore cannot generalise from individual learning to group learning. Consequently a continuation of conducting experiments in both settings is needed.

**Assessment/Feedback Space**

The learning process needs ratification through a testing regime. Pedagogical theories insist on the importance of feedback as a mechanism by which improvement can be made. The learning process requires a critical evaluative feedback loop. This will involve the provision of online questions and assessment in order to determine the status of the student's learning and the attainment which has been reached. Without this necessary feedback space it would be impossible to gauge whether learning has in fact taken place.

**Supervisor Space**

The dual roles of teacher and learner need to be reflected in the construction of a CSCL environment. Tutors would require their own private area for their specific tasks. It is suggested, following Kester *et al* [2006] that any interface that is constructed to assist collaborative research needs to ensure that supportive information and schematic information are presented at separate times.

Although it could be argued that these spaces might be required for good working and not just learning it is contended that these spaces are more essential to the process of learning than they are to just working. Working can take place without the need for these spaces though it is accepted that their inclusion may enhance the working process. Since work can take place without reflection, socialisation, assessment and tutorials these spaces distinguish the CSCL environment.

The results of all of these studies have their place in a consideration of collaborative domains and it will be important to take these results into account when defining CSCR. These together with the results of iterative user analyses will form the construction basis of a CSCR related interface or instance.

## 4.3 Identifying the Gaps between CSCW, CSCL and CSCR, and Determining CSCR Domain

This review has shown that there is no fully defined environment which meets all the needs of a research community. A series of gaps have been identified and the requirements will be examined now. So far we have looked at the established domains of CSCW and CSCL. This approach has brought a more rigorous definition and distinction to each of these domains in that they are shown to be related to each other where CSCL is a specialised type of CSCW and all of the features of CSCW are contained in CSCL.

However the literature examination has shown that these domains are insufficient to provide a rich enough environment for collaborative research. A number of additional areas are required in order to fill in the gaps left by the CSCW and CSCL domains. The additional requirements needed by collaborative research will now be examined

There are a number of differences between CSCR and CSCL which include the need to cater for the specialist requirements of research which has been defined as the acquisition of new knowledge. This includes such things as a complete record of all interactions between participants, which is an important and necessary tool to evaluate the contributions of each member in a collaboration group that can later on determine "a fair capital share" if the undergoing research project is successful. This is more relevant to collaboration between partners in different institutions where the division of funding maybe dependant upon contributed weighting. This would be contained in what may be called a "Knowledge Space". In addition there will be further requirements for a private space, public space, publication space and negotiation space to construct a CSCR domain:. Each of these will now be considered in turn.

### Knowledge Space

Research collaboration will generate its own knowledge base and a database system will be required which can store and retrieve this information as well as allocating ownership to individual contributions to ensure an appropriate apportionment of credit. It would be expected that this system would incorporate hypertext and links to bring cohesion to individual contributions, which is a form of cross referencing. Knowledge space is a repository which can track individual contributions of researchers and which will hold the data that will eventually feed into publication space for the construction of work to publishable standards.

### Private Space

Research is commonly the domain of groups of workers rather than individuals though not exclusively so. Each research group will need to have its own private area in which to work that is closed to non-group members. Since the knowledge is new knowledge primacy of publication becomes important and confidentiality is therefore essential to this process. It is important to maintain a secure area where work is developed before it is published.

### Public Space

The collaborative research group may wish to provide information upon the nature of the research which is being done, to encourage contributions, questions, raise issues etc. which can be placed online in the public domain. (e.g. online questionnaires, public bulletin boards etc). Public notification is important to engender contributions from outside the research group which may prove valuable both as a spur to new ideas and a source of research data itself.

### Negotiation Space

It is also clear that a CSCR domain will require space for negotiation between collaborators in order to enable free and frank discussion and to eliminate disagreements. Swaab et al [2004] have concluded that negotiation support systems should stimulate a common cultural identity among the individual participants and negotiation support systems should provide information to develop shared cognition among negotiators. Negotiation space will therefore be an important part of the definition of CSCR.

Group research may often introduce conflicts of opinion which need to be worked through on-line. This is more difficult online and may involve intensive and protracted discussions. This could be done by chat, forum or recorded video conferencing. It is envisaged that a CSCR domain may require a negotiation support system as discussed by Swaab *et al* [2004] in order to foster the resolution of possible conflicts arising between research collaborators. Conflicts between collaborators can cause unwanted stress generated in collaborative environments [Lawless and Allan 2004]. The provision of negotiation space is included in the CSCR domain to provide a mechanism for relieving stress in an on line collaborative scenario and by a careful management of the working processes.

### Publication Space

The ultimate aim of research is to provide to new knowledge to the research community. This is normally done through the mechanism of publication and as such is a vital and necessary part of the research process itself. The need for assistance afforded to the publication process should be incorporated the CSCR domain. This may include the provision of schemas templates specific journal style sheets as well as more application centred assistance in the form of a collaborative composition and publishing system such as CAWS [Liccardi et al 2007] The publication of pre-prints, e-prints and draft papers to online sites such as arxiv.org would be assisted by an automated process incorporated into the system.

**Additional Features**

Additional issues have been raised by other researchers such as the coupling of work and its organisation, informal communication and informal roles, awareness in distributed design, establishment of common grounds and perspective, clarification and convergence mechanisms in co-design. D'etienne [2006] These have been assessed as not of primary significance in the determination of the CSCR domain.

The full delineation of the differences and interdependencies of CSCW, CSCL and CSCR are summarised in table 3

Table 3: Degrees of Collaborative Space

| The Spaces required by each of the collaborative areas | | |
|---|---|---|
| **CSCW WorkingSpace** | **CSCL LearningSpace** | **CSCR ResearchSpace** |
| Administration | Administration | Administration |
| Communication | Communication | Communication |
| Scheduling | Scheduling | Scheduling |
| Sharing | Sharing | Sharing |
| Product | Product | Product |
|  | Reflection | Reflection |
|  | Social | Social |
|  | Assessment/Feedback | Assessment/Feedback |
|  | Supervisor | Supervisor |
|  |  | New Knowledge |
|  |  | Privacy |
|  |  | Public |
|  |  | Negotiation |
|  |  | Publication |

## 5. COMPARISON WITH VRE ENVIRONMENTS

The CSCR domain is not a Virtual Research Environment (VRE). The CSCR domain being proposed here is distinct from other environments in a number of key ways. For instance CSCR focuses on people, while VRE focuses on tools. CSCR specifies the necessary and sufficient conditions for setting up a valid collaborative research environment. CSCR is a logical domain while a VRE is a possible instance of that domain.

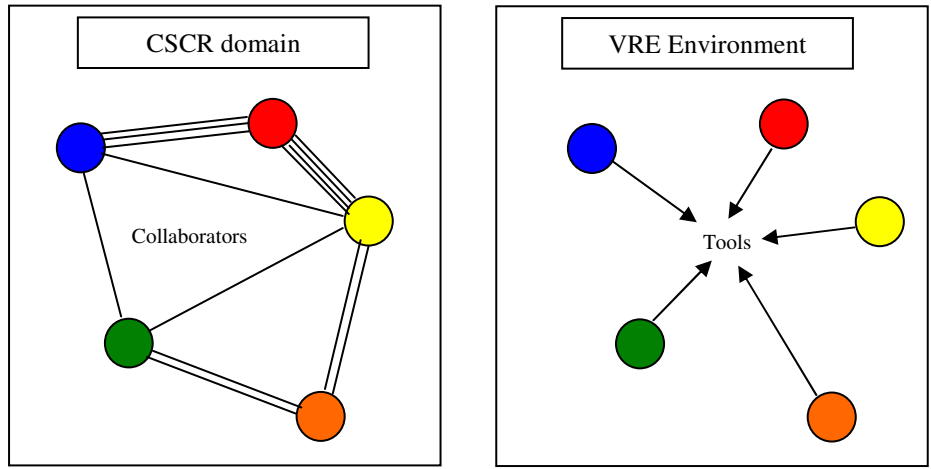

Figure 2 Comparison of Research Environments

The VRE has a range of tools necessary for researchers to be supported in their activities but it does not necessarily support collaborative activities. VREs such as that discussed in Wills [2005] concentrate on the structures needed to support individual roles, rather than collaborative ones.

The CSCR domain may act as a container for the VRE as well as a range of other tools. As such it is a domain using a portal to bring the focus upon collaborative research.

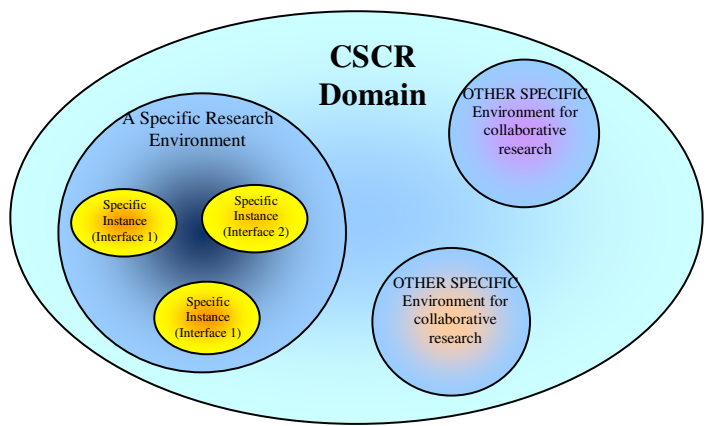

Figure 3 Relationship between domains, environments and Interfaces

Figure 3 shows the relationship between domains, environments and interfaces. The domain is defined by a set of 14 specific spaces. See Table 3. The environment will be defined by a specific set of tools, and the interface will be defined by a specific arrangement of these tools in a portal framework.

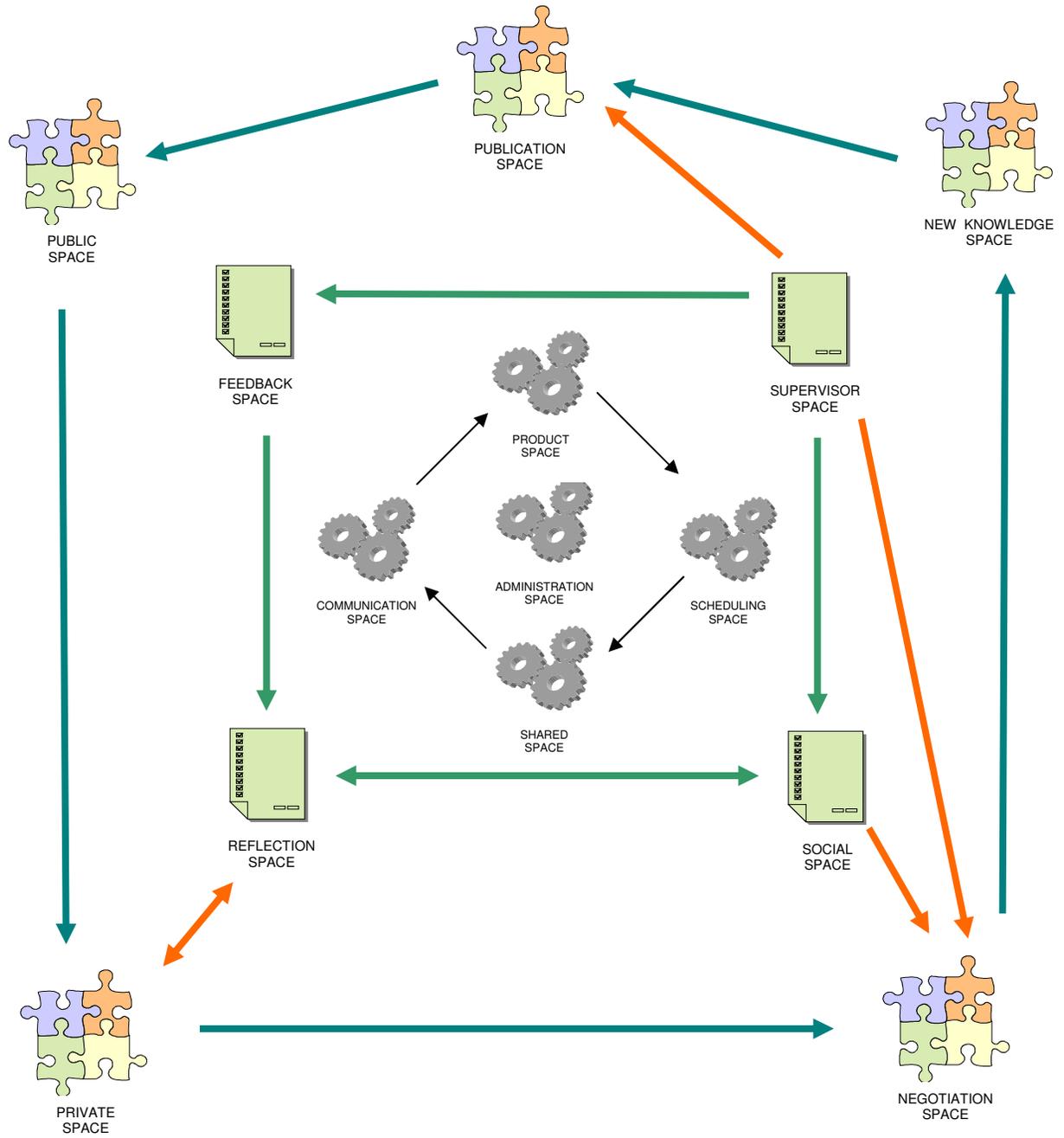

Figure 4: The three domains CSCW,CSCL and CSCR process representation

## 6 SUMMARY

It has been argued that there is a case to be made for regarding CSCR as a separate and distinct area of investigation. Each of these domains CSCW, CSCL and CSCR has their own specification and requirements. The first two according to Stahl, G. [2003] are defined by having their own *"conferences, journals and adherence."* The latter is yet to develop and is an emerging area of research.

All three domains, CSCW, CSCL and CSCR have a commonality, with CSCL and CSCR having dependency on CSCW but CSCR has individual aspects which are not part of the other two, (see figure 4) and consequently is distinct and should be treated as such. Hinze-Hoare [2006c].

There are now numerous examples of these VREs and research environments. Newly developing on-line scientific web-logs and other portals which enable scientists around the world to perform an on-line collaboration over the internet are being created, Handoko [2005], but with little thought as to the specification, definition or requirements of such a domain.

The definition of the CSCR domain in this paper is presented as universally applicable and determining for all potential collaborative research environments.

A following paper will take this analysis further by considering which tools are necessary to determine a particular instance of the CSCR domain. Application will be made to the construction of a Collaborative Research Environment for Students and Supervisors (CRESS).